\documentclass[prl,twocolumn,letterpaper, superscriptaddress]{revtex4-1}
\usepackage{graphicx}
\usepackage{dcolumn}
\usepackage{bm}
\usepackage{color}
\usepackage{tabularx}
\usepackage{array}
\usepackage{amsmath}
\usepackage{stmaryrd}  

\begin{document}

\title{Selective control of vortex polarities by microwave field in two robustly synchronized spin-torque nano-oscillators}

\author{Yi Li}
\email{yi.li@cea.fr}
\affiliation{Service de Physique de l'\'{E}tat Condens\'{e}, CEA, CNRS, Universit\'{e} Paris-Saclay, Gif-sur-Yvette, France}

\author{Xavier de Milly}
\affiliation{Service de Physique de l'\'{E}tat Condens\'{e}, CEA, CNRS, Universit\'{e} Paris-Saclay, Gif-sur-Yvette, France}

\author{Olivier Klein}
\affiliation{SPINTEC, Univ. Grenoble Alpes / CEA / CNRS, 38000 Grenoble, France}

\author{Vincent Cros}
\affiliation{Unit\'{e} Mixte de Physique CNRS, Thales, Univ. Paris-Sud, Universit\'{e} Paris-Saclay, Palaiseau, France}

\author{Julie Grollier}
\affiliation{Unit\'{e} Mixte de Physique CNRS, Thales, Univ. Paris-Sud, Universit\'{e} Paris-Saclay, Palaiseau, France}

\author{Gr\'{e}goire de Loubens}
\email{gregoire.deloubens@cea.fr}
\affiliation{Service de Physique de l'\'{E}tat Condens\'{e}, CEA, CNRS, Universit\'{e} Paris-Saclay, Gif-sur-Yvette, France}

\date{\today}

\begin{abstract}

Manipulating operation states of coupled spin-torque nano-oscillators (STNOs), including their synchronization, is essential for applications such as complex oscillator networks. In this work we experimentally demonstrate selective control of two coupled vortex STNOs through microwave-assisted switching of their vortex core polarities. First, the two oscillators are shown to synchronize due to dipolar interaction in a broad frequency range tuned by external biasing field. Coherent output is demonstrated along with strong linewidth reduction. Then, we show individual vortex polarity control of each oscillator, which leads to synchronization/desynchronization due to accompanied frequency shift. Our methods can be easily extended to multiple-element coupled oscillator networks.

\end{abstract}

\maketitle

\indent A growing interest has been witnessed on spin-torque nano-oscillators (STNOs) \cite{KiselevNature2003}. They exhibit numerous advantages in applications of modern electronics, such as nano-scale geometry, microwave-frequency signal output and frequency tunability. Particularly, STNOs can be coupled to each other leading to synchronization \cite{KakaNature2005,MancoffNature2005,RuotoloNnano2009,SaniNcomm2013,locatelliSREP2015,HoushangNnano2015,LebrunNComm2017}. This provides a platform to study synchronization phenomena \cite{PikovskyBook2001} as well as to mimic neural networks \cite{GrollierProcIEEE2016}. A more advanced utilization of STNO networks requires manipulation of synchronization states, or the ability to turn on or off synchronization. To this aim, regular methods include tuning the biasing current and magnetic field in order to vary the output frequencies. However they usually also lead to a modification of the device properties, which thus adds more complexity to the system.\\
\indent The introduction of vortex-based STNOs \cite{PribiagNphys2007,DussauxNcomm2010,LocatelliAPL2011} provides an alternative solution. They have been shown to synchronize efficiently with various coupling mechanisms \cite{SaniNcomm2013,locatelliSREP2015,LiPRL2017,LebrunNComm2017}. With the additional parameter of vortex core polarity, defined as the binary perpendicular direction of the magnetization at the vortex core \cite{ShinjoScience2000}, the sign of the frequency tunability \cite{deLoubensPRL2009} for each STNO can be independently adjusted by dynamically switching the corresponding vortex polarity \cite{VanWaeyenbergeNature2006,YamadaNmat2007,PigeauNphys2011}, without changing the properties of other devices. This leads to a simple demonstration and control of synchronization \cite{locatelliSREP2015,LiPRL2017}. Moreover, the inter-device coupling strength can be also changed from their relative polarity alignments \cite{AbreuAraujoPRB2015,locatelliSREP2015,LiPRL2017}, which provides a new method to modify STNO networks.\\
\indent In this work we explore microwave-assisted vortex polarity switching in a system composed of two dipolarly coupled STNOs. First, by further reducing the inter-device spacing of the two STNOs down to 50 nm compared with our previous work \cite{locatelliSREP2015}, we show their improved synchronization in a broad frequency range with coherent power output and linewidth reduction due to stronger dipolar coupling \cite{BelanovskyPRB2012,BelanovskyAPL2013}. Then, we show that in the synchronized state, a microwave field of well chosen frequency and amplitude can selectively switch the vortex polarity of each STNO, leading to two distinct regions of reversal events in the switching portrait. Owing to a nonzero biasing field, the polarity switching is accompanied by a large shift of the auto-oscillation frequency of the corresponding STNO, which results in the turn-off of the synchronization. Furthermore our measurements of the switching portraits in the unsynchronized states yield nearly identical regions for core reversal, which indicates that the switching conditions of two strongly interacting STNOs are mostly insensitive to their coupling. These results provide a new way to control coupled vortex-based STNO arrays.\\
\begin{figure}[htb]
 \centering
 \includegraphics[width=3.2 in]{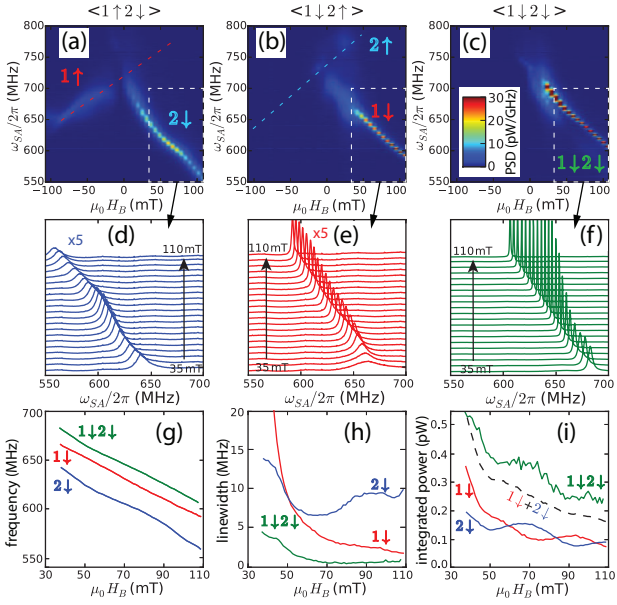}
 \caption{
 (a-c) Power spectral densities of (a) $\langle1\uparrow 2\downarrow\rangle$, (b) $\langle1\downarrow 2\uparrow\rangle$ and (c) $\langle1\downarrow 2\downarrow\rangle$ polarity states as a function of $\mu_0H_B$. (d-f) Zoomed-in spectral lineshapes of (a-c) in the white box regions, from 35 mT to 110 mT. (g-i) Field dependence of (g) frequency, (h) linewidth and (i) integrated powers extracted from (d-f). In (i) the black dashed curve shows the power sum of red and blue curves.}
 \label{fig1}
\end{figure}
\indent The sample consists of two adjacent cylindrical spin-valve nanopillars with layer structure of Py(15 nm)/Cu(10 nm)/Py(4 nm) (Py = Ni$_{80}$Fe$_{20}$). They have identical nominal diameters of 200 nm and an edge-to-edge separation of 50 nm. An antenna is fabricated on top of the sample to generate an in-plane microwave field $h_{rf}$. During the experiments the common dc current is set to 41 mA which is twice the critical current for spin transfer induced vortex core auto-oscillation. The auto-oscillation is dominated by the Py(15 nm) layers, whose polarity states will be referred to throughout the paper. The thinner Py layers act as the polarizers \cite{KhvalkovskiyAPL2010}. Details of the vortex STNO pair operations can be found elsewhere \cite{locatelliSREP2015,LiPRL2017}. A perpendicular biasing field $H_B$ is applied to tune the output frequency \cite{deLoubensPRL2009,LocatelliAPL2011}. \\
\indent First we demonstrate mutual synchronization of the two STNOs, labeled as 1 and 2. Figs. \ref{fig1}(a-c) show the power spectral density (PSD) as a function of frequency $\omega_{SA}/2\pi$ for various $H_B$. Three polarity states, $\langle1\uparrow 2\downarrow\rangle$, $\langle1\downarrow 2\uparrow\rangle$ and $\langle1\downarrow 2\downarrow\rangle$, have been studied which are set by microwave-assisted polarity switching as will be discussed later. The $\langle\uparrow\rangle$ (or $\langle\downarrow\rangle$) state is defined as the polarity pointing towards (or away from) the positive $H_B$ direction. For each antiparallel polarity states (Figs. \ref{fig1}a,b), two branches of auto-oscillation signals can be observed, with opposite frequency dependence on $H_B$ due to different polarity alignments to field \cite{deLoubensPRL2009,jenkinsAPL2014}. We focus on the spectra of $\langle1\downarrow\rangle$ and $\langle2\downarrow\rangle$ STNOs in the large positive-field region, with the lineshapes plotted in Figs. \ref{fig1}(d-e). Their output frequencies are well separated ($>100$ MHz) from their $\langle\uparrow\rangle$ neighbors (see extrapolations of the faint branches in Figs. \ref{fig1}(a) and (b) shown by dashed red and blue lines, respectively). Thus the influence of inter-device dipolar coupling is negligible and the two spectra can be taken as the individual outputs of the STNOs 1 and 2 with $\langle\downarrow\rangle$ polarity states. \\
\indent Then we move to the parallel polarity state $\langle1\downarrow 2\downarrow\rangle$. Instead of two peaks corresponding to the superposition of the $\langle1\downarrow\rangle$ and $\langle2\downarrow\rangle$ spectra, only one auto-oscillation peak is observed in Fig. \ref{fig1}(f), with a much stronger amplitude and smaller linewidth in a broad frequency (biasing field) range. The frequencies, full-width half-maximum linewidths and integrated powers are extracted and plotted in Figs. \ref{fig1}(g-i), respectively. For the $\langle1\downarrow 2\downarrow\rangle$ state, the linewidth is greatly reduced from the two individual peaks (Fig. \ref{fig1}h) by more than a factor of two in the entire field range; the output power, plotted with the green curve in Fig. \ref{fig1}(i), is larger than the sum of the two individual devices marked by the black dashed curve. All those evidences show that the two STNOs are mutually synchronized, with coherent output and reduced phase noise. It agrees with our previous observations of mutual synchronization due to dipolar coupling \cite{locatelliSREP2015,LiPRL2017}. \\
\begin{figure}[htb]
 \centering
 \includegraphics[width=3.0 in]{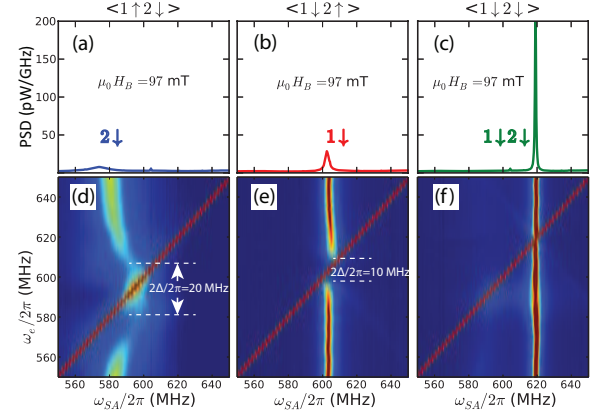}
 \caption{(a-c) Power spectral densities of (a) $\langle1\uparrow 2\downarrow\rangle$, (b) $\langle1\downarrow 2\uparrow\rangle$ and (c) $\langle1\downarrow 2\downarrow\rangle$ states at $\mu_0H_B=97$ mT. (d-f) Power spectral density of (a-c) in the presence of a microwave field ($P_e=-16$ dBm) at various frequencies $\omega_e/2\pi$.}
 \label{fig2}
\end{figure}
\indent Next we examine the vulnerability of the synchronization state to external perturbation, a weak microwave field in our case. For the three polarity states in Fig. \ref{fig1}, we set the biasing field to $\mu_0H_B = 97$ mT where clear auto-oscillation spectra can be observed (Figs. \ref{fig2}a-c). Then a microwave power of $P_e=-16$ dBm is applied, which corresponds to a linear amplitude of $\mu_0h_{rf}= 0.4$ mT. By sweeping its frequency $\omega_e/2\pi$, phase-locking to the microwave field is observed \cite{RippardPRL2005,HamadehAPL2014} for $\langle1\downarrow\rangle$ (Figs. \ref{fig2}d) and $\langle2\downarrow\rangle$ (Figs. \ref{fig2}e) peaks with locking bandwidth $\Delta_e/2\pi$ of 10 MHz and 5 MHz, respectively. In contrast, the $\langle1\downarrow2\downarrow\rangle$ peak is barely influenced by the microwave field and the phase-locking bandwidth is negligible (Fig. \ref{fig2}f). In fact its resilience to the external microwave field is intrinsically connected to the much smaller output linewidth. Owing to the coherent emission, the synchronized two-STNO system is more resistant to environmental noise which is the main source of linewidth broadening \cite{JVKimPRL2008,GeorgesPRL2008,SlavinIEEE2009}. The observations in Fig. \ref{fig2} prove that dipolar-field-induced synchronization can improve the noise stability of STNOs. \\
\begin{figure}[htb]
 \centering
 \includegraphics[width=3.0 in]{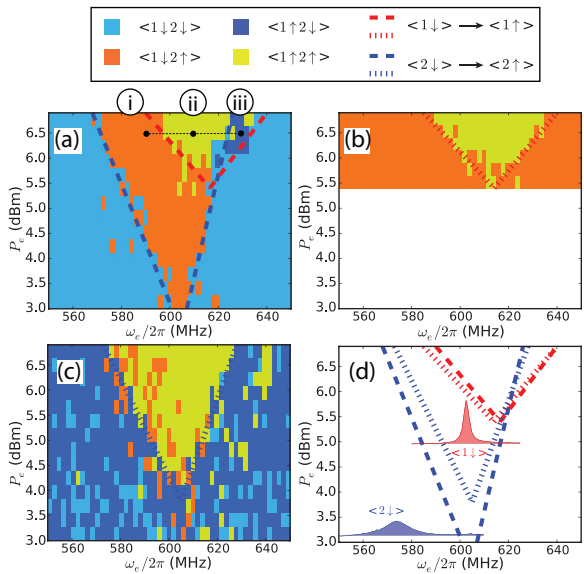}
 \caption{Microwave switching portraits of vortex polarities for (a) $\langle1\downarrow 2\downarrow\rangle$, (b) $\langle1\downarrow 2\uparrow\rangle$ and (c) $\langle1\uparrow 2\downarrow\rangle$ initial polarity states as demonstrated in Fig. \ref{fig1}(a-c). The bias field is set to $\mu_0H_B=97$ mT. The power range is 3.0 to 6.9 dBm in (a), (c) and 5.4 to 6.9 dBm in (b). The power steps are 0.3 dBm and the frequency steps are 2 MHz. The dashed and dotted lines define the boundary of polarity switching events. (d) Switching boundaries plotted in one figure. The blue and red cones correspond to the switching of STNO 1 and 2, respectively. The two lineshapes show the auto-oscillation peaks of $\langle1\downarrow\rangle$ and $\langle2\downarrow\rangle$, as measured in Fig. \ref{fig2}(a) and (b).}
 \label{fig3}
\end{figure}
\indent In the presence of a strong microwave field, however, the response of synchronization state is completely different from in Fig. \ref{fig2}(f), which provides the opportunity to address the vortex polarities independently. In Fig. \ref{fig3} we show our main results of microwave-assisted polarity switching for vortex auto-oscillators. This technique has already been demonstrated on a passive vortex \cite{VanWaeyenbergeNature2006,YamadaNmat2007,PigeauNphys2011}. For our samples, we firstly set the polarity state to $\langle1\downarrow 2\downarrow\rangle$ state with a strong negative field $\mu_0H_B=-300$ mT and move to $\mu_0H_B = 97$ mT, which favors the $\langle\uparrow\rangle$ states. In contract to Fig. \ref{fig2}, a much stronger microwave power $P_e$ is then applied through the antenna for 3 seconds in order to switch the vortex polarity, and the final state is read through its associated emission spectrum (e.g. Figs. \ref{fig2}a-c). Fig. \ref{fig3}(a) shows the switching results as a function of $P_e$ and $\omega_e/2\pi$. Four different colors represent all possible polarity states. For example, at $P_e=6.5$ dBm using $\omega_e/2\pi=590$, $610$ and $630$ MHz will set the system to (i) $\langle1\downarrow 2\uparrow\rangle$, (ii) $\langle1\uparrow 2\uparrow\rangle$ and (iii) $\langle1\uparrow 2\downarrow\rangle$ final states, respectively. The switching portrait can be categorized into two cone regions whose boundaries are marked by dashed lines: red for flipping $\langle1\downarrow\rangle$ to $\langle1\uparrow\rangle$ and blue for flipping $\langle2\downarrow\rangle$ to $\langle2\uparrow\rangle$. The cone-shaped switching boundary agrees well with previous reports on the polarity switching of a passive vortex \cite{PigeauNphys2011}. \\
\indent To study the relation between the two cone regions and the two STNOs, we repeat the switching experiments for different initial polarity states where oscillators are not synchronized, as $\langle1\downarrow 2\uparrow\rangle$ in Fig. \ref{fig3}(b) and $\langle1\uparrow 2\downarrow\rangle$ in Fig. \ref{fig3}(c). The noise in Fig. \ref{fig3}(c) is mainly due to $\langle1\uparrow 2\downarrow\rangle$ initialization errors. The boundaries of the new switching portraits are depicted by the dotted lines, which are plotted in Fig. \ref{fig3}(d) together with the dashed lines in Fig. \ref{fig3}(a). Due to the absence of synchronization, the switching condition of $\langle1\downarrow\rangle$ and $\langle2\downarrow\rangle$ should be different from measured in Fig. \ref{fig3}(a). Surprisingly, they coincide pretty well. For comparison the output lineshapes of $\langle1\downarrow\rangle$ and $\langle2\downarrow\rangle$ are also plotted in Fig. \ref{fig3}(d). The optimal switching frequencies at the bottom of the cone regions are shifted to slightly higher values than the corresponding auto-oscillation peaks, by 13 MHz for $\langle1\downarrow\rangle$ and 29 MHz for $\langle2\downarrow\rangle$. \\
\indent For two interacting vortices, usually their dynamics cannot be disentangled. The polarity states of a coupled vortex pair have been manipulated by resonant microwave excitations, which can only determine the relative polarity state without addressing the state of individual vortices \cite{JainNcomm2012,JainAPL2014, HanzeSREP2016}. In this scenario, the microwave field acts most effectively at the frequency of the hybridized modes. In the case of two coupled \emph{auto-oscillating} vortices, however, the most effective switching frequency is independent of their mutual synchronization dynamics. Instead the switching pattern in Fig. \ref{fig3}(a) almost equals to the superposition of Figs. \ref{fig3}(b) and (c). The observation can be explained by the different nature of the phase equation for an auto-oscillator, in which the dipolar inter-device coupling \emph{competes} with microwave-oscillator coupling instead of adapting to it. Our experiments indicate that at the threshold of polarity switching, the dipolar coupling is overwhelmed by the stronger microwave coupling. We justify this argument by comparing the two coupling strengths at the switching threshold. For switching powers which are 20 dB larger than those used in Figs. \ref{fig2}(a-b), the microwave coupling strength $\Delta/2\pi$ is in principle ten times greater \cite{HamadehAPL2014}, which is around 50 MHz for STNO 1 and 100 MHz for STNO 2. On the other hand, the dipolar coupling $\Omega/2\pi$ is estimated to be around 10 MHz in the parallel polarity state \cite{AbreuAraujoPRB2015,LiPRL2017} and is negligible compared with the microwave couplings. The upshift of the optimal switching frequency from the auto-oscillation frequency is due to the nonlinear frequency adjustment as the vortex gyration amplitude increases \cite{DussauxPRB2012}.\\
\indent Lastly we compare the switching power with the threshold value of a single vortex core. The latter is well understood, where the gyrating vortex core reaches a critical amplitude (velocity) at which its spatial deformation reaches a dynamical instability \cite{HertelPRL2007,GuslienkoPRL2008,LeePRL2008}. For Py, the critical velocity \cite{LeePRL2008} is about 320 m/s for a exchange stiffness of 12 pJ/m \cite{LiPRL2016}. At resonance, using the Thiele equation to calculate the gyration amplitude \cite{GuslienkoAPL2006}, we obtain the vortex core velocity as $v = \gamma\mu_0h_{rf}R/(3\sqrt{2}\alpha\eta)$, where $R=100$ nm is the radius of the nanodisc, $\alpha$ is the Gilbert damping of Py, and $\eta \sim 1.7$ is the topological renormalization of the damping \cite{DussauxPRB2012}. The factor $\sqrt{2}$ accounts for that the microwave field is linearly polarized. Using $\alpha=0.008$ for Py, the critical microwave field for vortex polarity switching is $\mu_0h_{rf}\sim 1$ mT, which is around 2 dBm for the geometry of our antenna. The measurements in Fig. \ref{fig3} yield similar switching powers for STNOs 1 and 2, which indicates that the injected dc current and corresponding spin transfer torque do not essentially change the behavior of a single vortex at large microwave drives. \\
\indent In summary, we have experimentally demonstrated coherent and robust synchronization of two vortex STNOs coupled by their dipolar interaction in a broad frequency range. Their synchronization state can be controlled by microwave-assisted vortex polarity switching. We highlight this device-selective, coupling-insensitive and channel-sharing polarity switching technique, which is a technical requirement in small and densely packed STNO networks. We also note that energy-efficient switching is possible with microwave pulses \cite{VanWaeyenbergeNature2006,PigeauNphys2011} for the operation of such networks. Our results add new understanding to the interacting mechanism of a strong microwave to two coupled auto-oscillators and provide new potential to dipolarly-coupled vortex STNOs for the implementation of coupled oscillator networks.
\indent We thank S. Giraud and C. Deranlot for their helps on sample growth and nanofabrication. We acknowledge the MEMOS project ANR-14-CE26-0021 and the MOSAIC project ICT-FP7 317950 for financial support.

\bibliographystyle{ieeetr}

\begin{thebibliography}{10}

\bibitem{KiselevNature2003}
S.~I. Kiselev, J.~C. Sankey, I.~N. Krivorotov, N.~C. Emley, R.~J. Schoelkopf,
  R.~A. Buhrman, and D.~C. Ralph, ``Microwave oscillations of a nanomagnet
  driven by a spin-polarized current,'' {\em Nature}, vol.~425, p.~380, 2003.

\bibitem{KakaNature2005}
S.~Kaka, M.~R. Pufall, W.~H. Rippard, T.~J. Silva, S.~E. Russek, and J.~A.
  Katine, ``Mutual phase-locking of microwave spin torque nano-oscillators,''
  {\em Nature}, vol.~437, p.~389, 2005.

\bibitem{MancoffNature2005}
F.~B. Mancoff, N.~D. Rizzo, B.~N. Engel, and S.~Tehrani, ``Phase-locking in
  double-point-contact spin-transfer devices,'' {\em Nature}, vol.~437, p.~393,
  2005.

\bibitem{RuotoloNnano2009}
A.~Ruotolo, V.~Cros, B.~Georges, A.~Dussaux, J.~Grollier, C.~Deranlot,
  R.~Guillemet, K.~Bouzehouane, S.~Fusil, and A.~Fert, ``Phase-locking of
  magnetic vortices mediated by antivortices,'' {\em Nature Nano.}, vol.~4,
  p.~528, 2009.

\bibitem{SaniNcomm2013}
S.~Sani, J.~Persson, S.~M. Mohseni, Y.~Pogoryelov, P.~K. Muduli, A.~Eklund,
  G.~Malm, M.~K\"{a}ll, A.~Dmitriev, and J.~\r{A}kerman, ``Mutually
  synchronized bottom-up multi-nanocontact spin-torque oscillators,'' {\em Nat.
  Commun.}, vol.~4, p.~2731, 2013.

\bibitem{locatelliSREP2015}
N.~Locatelli, A.~Hamadeh, F.~Abreu~Araujo, A.~D. Belanovsky, P.~N. Skirdkov,
  R.~Lebrun, V.~V. Naletov, K.~A. Zvezdin, M.~Mu\~{n}oz, J.~Grollier, O.~Klein,
  V.~Cros, and G.~de~Loubens, ``Efficient synchronization of dipolarly coupled
  vortex-based spin transfer nano-oscillators,'' {\em Sci. Rep.}, vol.~5,
  p.~17039, 2015.

\bibitem{HoushangNnano2015}
A.~Houshang, E.~Iacocca, P.~D\"{u}rrenfeld, S.~R. Sani, J.~\r{A}kerman, and
  R.~K. Dumas, ``Spin-wave-beam driven synchronization of nanocontact
  spin-torque oscillators,'' {\em Nat. Nano.}, vol.~11, p.~280, 2015.

\bibitem{LebrunNComm2017}
R.~Lebrun, S.~Tsunegi, P.~Bortolotti, H.~Kubota, A.~S. Jenkins, M.~Romera,
  K.~Yakushiji, A.~Fukushima, J.~Grollier, S.~Yuasa, and V.~Cros, ``Mutual
  synchronization of spin torque nano-oscillators through a non-local and
  tunable electrical coupling,'' {\em Nat. Commun.}, vol.~8, p.~15825, 2017.

\bibitem{PikovskyBook2001}
A.~Pikovsky, M.~Rosenblum, and J.~Kurths, {\em Synchronization: A universal
  concept in nonlinear sciences}.
\newblock Cambridge University Press, Cambridge, UK, 2001.

\bibitem{GrollierProcIEEE2016}
J.~Grollier, D.~Querlioz, and M.~D. Stiles, ``Spintronic nanodevices for
  bioinspired computing,'' {\em Proc. IEEE}, vol.~104, p.~2024, 2016.

\bibitem{PribiagNphys2007}
V.~S. Pribiag, I.~N. Krivorotov, G.~D. Fuchs, P.~M. Braganca, O.~Ozatay, J.~C.
  Sankey, D.~C. Ralph, and R.~A. Buhrman, ``Magnetic vortex oscillator driven
  by d.c. spin-polarized current,'' {\em Nature Phys.}, vol.~3, p.~498, 2007.

\bibitem{DussauxNcomm2010}
A.~Dussaux, B.~Georges, J.~Grollier, V.~Cros, A.~V. Khvalkovskiy, A.~Fukushima,
  M.~Konoto, H.~Kubota, K.~Yakushiji, S.~Yuasa, K.~A. Zvezdin, K.~Ando, and
  A.~Fert, ``Large microwave generation from current-driven magnetic vortex
  oscillators in magnetic tunnel junctions,'' {\em Nat. Commun.}, vol.~1, p.~8,
  2010.

\bibitem{LocatelliAPL2011}
N.~Locatelli, V.~V. Naletov, J.~Grollier, G.~de~Loubens, V.~Cros, C.~Deranlot,
  C.~Ulysse, G.~Faini, O.~Klein, and A.~Fert, ``Dynamics of two coupled
  vortices in a spin valve nanopillar excited by spin transfer torque,'' {\em
  Appl. Phys. Lett.}, vol.~98, p.~062501, 2011.

\bibitem{LiPRL2017}
Y.~Li, X.~de~Milly, F.~Abreu~Araujo, O.~Klein, V.~Cros, J.~Grollier, and
  G.~de~Loubens, ``Probing phase coupling between two spin-torque
  nano-oscillators with an external source,'' {\em Phys. Rev. Lett.}, vol.~118,
  p.~247202, Jun 2017.

\bibitem{ShinjoScience2000}
T.~Shinjo, T.~Okuno, R.~Hassdorf, K.~Shigeto, and T.~Ono, ``Magnetic vortex
  core observation in circular dots of permalloy,'' {\em Science}, vol.~289,
  p.~930, 2000.

\bibitem{deLoubensPRL2009}
G.~de~Loubens, A.~Riegler, B.~Pigeau, F.~Lochner, F.~Boust, K.~Y. Guslienko,
  H.~Hurdequint, L.~W. Molenkamp, G.~Schmidt, A.~N. Slavin, V.~S. Tiberkevich,
  N.~Vukadinovic, and O.~Klein, ``Bistability of vortex core dynamics in a
  single perpendicularly magnetized nanodisk,'' {\em Phys. Rev. Lett.},
  vol.~102, p.~177602, 2009.

\bibitem{VanWaeyenbergeNature2006}
B.~Van~Waeyenberge, A.~Puzic, H.~Stoll, K.~W. Chou, T.~Tyliszczak, H.~R.,
  M.~F\"{a}hnle, H.~Br\"{u}ckl, K.~. Rott, G.~Reiss, I.~Neudecker, D.~Weiss,
  C.~H. Back, and G.~Sch\"{u}tz, ``Magnetic vortex core reversal by excitation
  with short bursts of an alternating field,'' {\em Nature}, vol.~444, p.~461,
  2006.

\bibitem{YamadaNmat2007}
K.~Yamada, S.~Kasai, Y.~Nakatani, K.~Kobayashi, H.~Kohno, A.~Thiaville, and
  T.~Ono, ``Electrical switching of the vortex core in a magnetic disk,'' {\em
  Nature Mater.}, vol.~6, p.~270, 2007.

\bibitem{PigeauNphys2011}
B.~Pigeau, G.~de~Loubens, O.~Klein, A.~Riegler, F.~Lochner, G.~Schmidt, and
  L.~W. Molenkamp, ``Optimal control of vortex-core polarity by resonant
  microwave pulses,'' {\em Nature Phys.}, vol.~7, p.~26, 2011.

\bibitem{AbreuAraujoPRB2015}
F.~Abreu~Araujo, A.~D. Belanovsky, P.~N. Skirdkov, K.~A. Zvezdin, A.~K.
  Zvezdin, N.~Locatelli, R.~Lebrun, J.~Grollier, V.~Cros, G.~de~Loubens, and
  O.~Klein, ``Optimizing magnetodipolar interactions for synchronizing vortex
  based spin-torque nano-oscillators,'' {\em Phys. Rev. B}, vol.~92, p.~045419,
  2015.

\bibitem{BelanovskyPRB2012}
A.~D. Belanovsky, N.~Locatelli, P.~N. Skirdkov, F.~Abreu~Araujo, J.~Grollier,
  K.~A. Zvezdin, V.~Cros, and A.~K. Zvezdin, ``Phase locking dynamics of
  dipolarly coupled vortex-based spin transfer oscillators,'' {\em Phys. Rev.
  B}, vol.~85, p.~100409(R), 2012.

\bibitem{BelanovskyAPL2013}
A.~D. Belanovsky, N.~Locatelli, P.~N. Skirdkov, F.~Abreu~Araujo, K.~A. Zvezdin,
  J.~Grollier, V.~Cros, and A.~K. Zvezdin, ``Numerical and analytical
  investigation of the synchronization of dipolarly coupled vortex spin-torque
  nano-oscillators,'' {\em Appl. Phys. Lett.}, vol.~103, p.~122405, 2013.

\bibitem{KhvalkovskiyAPL2010}
A.~V. Khvalkovskiy, J.~Grollier, N.~Locatelli, Y.~V. Gorbunov, K.~A. Zvezdin,
  and V.~Cros, ``Nonuniformity of a planar polarizer for spin-transfer-induced
  vortex oscillations at zero field,'' {\em Appl. Phys. Lett.}, vol.~96,
  p.~212507, 2010.

\bibitem{jenkinsAPL2014}
A.~S. Jenkins, E.~Grimaldi, P.~Bortolotti, R.~Lebrun, H.~Kubota, K.~Yakushiji,
  A.~Fukushima, G.~de~Loubens, O.~Klein, S.~Yuasa, and V.~Cros, ``Controlling
  the chirality and polarity of vortices in magnetic tunnel junctions,'' {\em
  Appl. Phys. Lett.}, vol.~105, p.~172403, 2014.

\bibitem{RippardPRL2005}
W.~H. Rippard, M.~R. Pufall, S.~Kaka, T.~J. Silva, S.~E. Russek, and J.~A.
  Katine, ``Injection locking and phase control of spin transfer
  nano-oscillators,'' {\em Phys. Rev. Lett.}, vol.~95, p.~067203, 2005.

\bibitem{HamadehAPL2014}
A.~Hamadeh, N.~Locatelli, V.~V. Naletov, R.~Lebrun, G.~de~Loubens, J.~Grollier,
  O.~Klein, and V.~Cros, ``Perfect and robust phase-locking of a spin transfer
  vortex nano-oscillator to an external microwave source,'' {\em Appl. Phys.
  Lett.}, vol.~104, p.~022408, 2014.

\bibitem{JVKimPRL2008}
J.-V. Kim, V.~Tiberkevich, and A.~N. Slavin, ``Generation linewidth of an
  auto-oscillator with a nonlinear frequency shift: Spin-torque
  nano-oscillator,'' {\em Phys. Rev. Lett.}, vol.~100, p.~017207, Jan 2008.

\bibitem{GeorgesPRL2008}
B.~Georges, J.~Grollier, M.~Darques, V.~Cros, C.~Deranlot, B.~Marcilhac,
  G.~Faini, and A.~Fert, ``Coupling efficiency for phase locking of a spin
  transfer nano-oscillator to a microwave current,'' {\em Phys. Rev. Lett.},
  vol.~101, p.~017201, Jul 2008.

\bibitem{SlavinIEEE2009}
A.~Slavin and V.~Tiberkevich, ``Nonlinear auto-oscillator theory of microwave
  generation by spin-polarized current,'' {\em IEEE Trans. Magn.}, vol.~45,
  p.~1875, 2009.

\bibitem{JainNcomm2012}
S.~Jain, V.~Novosad, F.~Fradin, J.~Pearson, V.~Tiberkevich, A.~Slavin, and
  S.~Bader, ``From chaos to selective ordering of vortex cores in interacting
  mesomagnets,'' {\em Nat. Commun.}, vol.~3, p.~1330, 2012.

\bibitem{JainAPL2014}
S.~Jain, V.~Novosad, F.~Y. Fradin, J.~E. Pearson, and S.~D. Bader, ``Dynamics
  of coupled vortices in perpendicular field,'' {\em Appl. Phys. Lett.},
  vol.~104, p.~082409, 2014.

\bibitem{HanzeSREP2016}
M.~H\"{a}nze, C.~F. Adolff, B.~Schulte, J.~M\"{o}ller, M.~Weigand, and
  G.~Meier, ``Collective modes in three-dimensional magnonic vortex crystals,''
  {\em Sci. Rep.}, vol.~6, p.~22402, 2016.

\bibitem{DussauxPRB2012}
A.~Dussaux, A.~V. Khvalkovskiy, P.~Bortolotti, J.~Grollier, V.~Cros, and
  A.~Fert, ``Field dependence of spin-transfer-induced vortex dynamics in the
  nonlinear regime,'' {\em Phys. Rev. B}, vol.~86, p.~014402, 2012.

\bibitem{HertelPRL2007}
R.~Hertel, S.~Gliga, M.~F\"{a}hnle, and C.~M. Schneider, ``Ultrafast
  nanomagnetic toggle switching of vortex cores,'' {\em Phys. Rev. Lett.},
  vol.~98, p.~117201, 2007.

\bibitem{GuslienkoPRL2008}
K.~Y. Guslienko, K.-S. Lee, and S.-K. Kim, ``Dynamic origin of vortex core
  switching in soft magnetic nanodots,'' {\em Phys. Rev. Lett.}, vol.~100,
  p.~027203, Jan 2008.

\bibitem{LeePRL2008}
K.-S. Lee, S.-K. Kim, Y.-S. Yu, Y.-S. Choi, K.~Y. Guslienko, H.~Jung, and
  P.~Fischer, ``Universal criterion and phase diagram for switching a magnetic
  vortex core in soft magnetic nanodots,'' {\em Phys. Rev. Lett.}, vol.~101,
  p.~267206, Dec 2008.

\bibitem{LiPRL2016}
Y.~Li and W.~E. Bailey, ``Wave-number-dependent gilbert damping in metallic
  ferromagnets,'' {\em Phys. Rev. Lett.}, vol.~116, p.~117602, Mar 2016.

\bibitem{GuslienkoAPL2006}
K.~Y. Guslienko, ``Low-frequency vortex dynamic susceptibility and relaxation
  in mesoscopic ferromagnetic dots,'' {\em Appl. Phys. Lett.}, vol.~89,
  p.~022510, 2006.

\end{thebibliography}

\end{document}